\pdfoutput=1

\documentclass{article}
\usepackage{iclr2026_conference,times}

\usepackage{amsmath,amsfonts,bm}

\def\eqref#1{equation~\ref{#1}}

\def\1{\bm{1}}

\DeclareMathAlphabet{\mathsfit}{\encodingdefault}{\sfdefault}{m}{sl}
\SetMathAlphabet{\mathsfit}{bold}{\encodingdefault}{\sfdefault}{bx}{n}

\usepackage{float} 
\usepackage{hyperref}
\hypersetup{
    colorlinks = true,
    citecolor=cyan,
    linkcolor=blue,
}
\usepackage{url}
\usepackage{xcolor}
\usepackage{mathtools}
\usepackage{fancyvrb}
\VerbatimFootnotes
\usepackage{graphicx}
\graphicspath{{./}{arXiv/}}
\usepackage{subcaption}
\usepackage{comment}

\usepackage[dvipsnames]{xcolor}
\usepackage{array, colortbl}
\usepackage{caption}

\definecolor{PLGAcol}{RGB}{214,128,58}
\definecolor{PLLAPCLcol}{RGB}{120,170,96}
\definecolor{LATMCcol}{RGB}{104,134,206}
\definecolor{RuleColor}{gray}{0.82}

\title{

Discovery of Bioresorbable Polymer Suture Coatings for Controlled Tissue Regeneration with Multimodal Foundation Models 

}

\author{Raghav Ramji \\
\texttt{raghavramji@gmail.com}}

\iclrfinalcopy

\begin{document}

\maketitle

\begin{abstract}
Therapeutic sutures can localize bioactive payloads to wounds, but burst release and insufficient coating durability can compromise regeneration and increase infection risk. In this work, we introduce a data-driven discovery strategy for bioresorbable polymer suture coatings that improve therapeutic retention and release kinetics. We develop \textbf{GenPoly}, an end-to-end multimodal generative framework that learns a shared polymer representation over paired polymer string sequences and molecular graphs via contrastive alignment, and couples this representation to property prediction, synthetic accessibility filtering, and motif-based candidate clustering to discover coating families beyond commonly used linear aliphatic polyesters. Starting from a clinically prevalent baseline, GenPoly identifies candidate coatings and selects representative polymers for experimental evaluation on nanoparticle-loaded sutures  with recombinant human EGF to an in vitro human keratinocyte wound model. A discovered polymer coating of the LA-TMC family shows improved payload retention and sustained release, exhibiting a $3.3\times$ increase in rhEGF concentration, accompanied by more localized Ki-67–positive keratinocyte proliferation at the wound bed, in comparison with the current state-of-the-art candidates. These findings suggest new opportunities in AI-driven polymer discovery, and highlight the viability of GenPoly in engineering bioresorbable coatings for suture-based therapeutic delivery in targeted tissue regeneration. 

\end{abstract}

\section{Introduction}

Therapeutic delivery approaches for tissue regeneration and wound healing can suffer from suboptimal payload release dynamics. Unlike traditional drug delivery settings, burst release results in hyperconcentrated payload delivery to the target site, overwhelming the cell differentiation process. Simultaneously, surgical sutures carrying the therapeutic can lack sufficient mechanical integrity, losing shape and failing to achieve controlled release. These factors may complicate healing and increase in vivo susceptibility to bacterial infection. Polymeric suture topcoats have proven to be critical in influencing both the structural integrity of the scaffold as well as protection of therapeutic payloads. State-of-the-art implementations primarily consider poly(lactide-co-glycolide) acid (PLGA) and similar linear aliphatic polyesters such as poly(glycolic) acid (PGA), poly(lactic) acid (PLA) -- particularly in nanoparticle-mediated delivery and mRNA transfection. Their wide availability, biocompatibility, and biodegradability are key reasons for their prevalence. However, identifying \textit{durable} and \textit{ductile} polymer topcoats within the combinatorially large space of candidates is crucial to avoid the aforementioned risks and facilitate targeted and  accelerated wound healing.

In this work, we develop \textbf{GenPoly}, an end-to-end pipeline for polymer discovery, 
with an application to target 
suture coatings that are bioresorbable (biocompatible, biodegradable, and bioabsorbable), mechanically robust, and viable for nanoparticle delivery to the wound site. We first train a foundation model over polymer SELFIES \citep{savit-etal-2025-polybart} and graph representations, with Transformer and graph convolutional network (GCN) encoders, cross-modal latent space alignment, and a Transformer SELFIES decoder. We subsequently performed multi-task learning over 8 relevant polymer properties, and perform nearest-neighbors search to obtain a shortlist of candidates given a state-of-the-art polymer for the application setting. The candidates are then grouped and filtered using a retrieval-augmented generation (RAG) pipeline; representative polymers from the top clusters
were obtained for use as a silk suture topcoat. 
For an in vitro study of wound healing in human keratinocytes, a 
polyelectrolyte nanoparticle complex was synthesized with human recombinant EGF protein (henceforth, rhEGF) payload. Analysis of rhEGF expression through ELISA reveals that the suture coated with an identified polymer -- namely, of the poly(l-lactide-co-trimethylene carbonate) (LA-TMC) family -- retains 
the payload for targeted wound healing, exhibiting controlled release with an increase in concentration. Immunofluorescence analysis
provides corroborating qualitative evidence, with more targeted growth and facilitated cell differentiation at the wound bed for the LA-TMC-coated suture in contrast to the others. 
Overall,  our pipeline serves as a tool for effective polymer discovery, exemplified through developing suture coatings that facilitate controlled release of therapeutic payloads in tissue engineering, with potential for further applications. 

\begin{figure}[t]
    \centering
    \includegraphics[width=1.0\linewidth]{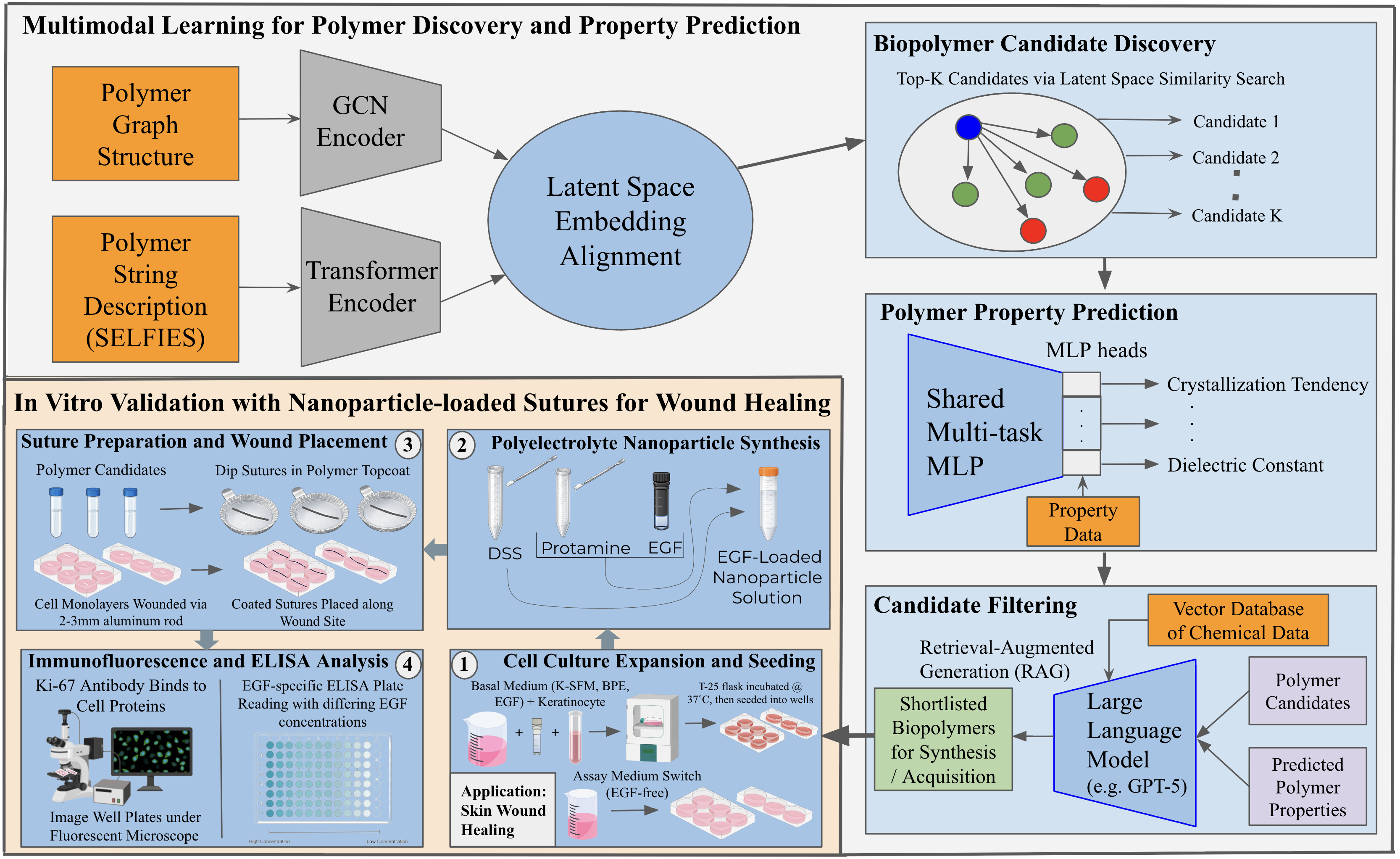}
    \caption{The complete GenPoly pipeline, from polymer data curation to multimodal model pre-training, property prediction and candidate generation, along with filtering and clustering to finalize candidates for wet lab testing. Subsequently, the candidates are used for in vitro wound healing with keratinocytes in suture-mediated therapeutic delivery via EGF-loaded nanoparticles. }
    \label{fig:main-pipeline}
\end{figure}

\section{Results}

\subsection{Method: Multimodal Learning for Polymer Discovery and Multi-task Property Prediction}\label{main:method}

\paragraph{Modalities and Architecture.}

We trained a multimodal model on polymer graph representations and Pseudo-polymer SELFIES (PSELFIES) strings \citep{savit-etal-2025-polybart}, following pre-processing using polymer SMILES \citep{Weininger1988SMILES} (PSMILES) sourced from the PI1M dataset \citep{Ma2020PI1M}. The PSMILES were canonicalized to molecular SMILES leveraging the RDKit package and converted to SELFIES \citep{krenn2020selfies}. They were then converted to PSELFIES by replacing the placeholder $[^*]$ atom with the rarely-occurring Astatine ([At]) atoms as the terminus of the chain, following \citet{savit-etal-2025-polybart}. Simultaneously, the PSMILES were converted to a molecular graph using node and edge information via RDKit and through the PyTorch Geometric library. This results in a dataset of 988,223 paired polymer string-graph representations for multimodal learning. 

Our architecture consists of three fundamental components, detailed in Appendix \ref{architecture}: (1) a GCN \citep{kipf2017semisupervised} encoder, (2) a Transformer \citep{vaswani2023attentionneed} encoder, and (3) a Transformer decoder. The polymer graph is encoded by a GCN, while in parallel, we train a Transformer encoder on tokenized PSELFIES strings. We obtain sequence embeddings with MLP projection heads for each, and learn cross-modal alignment via a CLIP-style \citep{radford2021learningtransferablevisualmodels} InfoNCE loss with all-pairs matrix $\ell$: 
\[\mathcal{L}_{InfoNCE} = -\frac{1}{2N}\sum_{i=1}^N\left[\log\frac{e^{\ell_{ii}}}{\sum_{j=1}^N e^{\ell_{ij}}} + \log\frac{e^{\ell_{ii}}}{\sum_{j=1}^N e^{\ell_{ji}}}  \right]\]

We then train a Transformer decoder to reconstruct the PSELFIES sequence by next token (symbol) prediction, with a variational autoencoder (VAE)-style token-level reconstruction loss $\mathcal{L} = -\frac{1}{M}\sum\limits_{t=1}^M \log p_{\theta}(y_t \mid y_{<t}, z)$.  The final hyperparameters and settings used to train our model are included in Appendix \ref{curves-hyperparameters}. Figures \ref{fig:contrastive} and \ref{fig:reconstruction} demonstrate stable training dynamics for the model in InfoNCE and reconstruction loss, suggesting that the decoder learns successful generation of PSELFIES, while the string and graph representations are well-aligned in the latent space.  

\paragraph{Polymer Property Prediction.}

To assess the suitability of polymers generated by our model in physical, electronic, and dielectric properties, we perform multi-task property prediction, following \citet{KUENNETH2021100238}. Values for 8 properties (described in Appendix \ref{property-prediction}) were sourced from the Khazana Polymer Informatics dataset, containing 6,264 entries obtained from Density Functional Theory (DFT) simulations. PSMILES were converted to a Morgan fingerprint via RDKit (embedding dim. = 2048), with a 3-layer shared MLP with property-specific heads (2-layer MLPs), with a multi-task mean squared error loss. Figure \ref{fig:prop-pred} shows that both train and test losses stabilize relatively quickly; the final hyperparameters are included in Appendix \ref{curves-hyperparameters}. 

\subsection{In Vitro Application: Wound Healing with Nanoparticle-loaded Sutures}

\paragraph{Experimental Setting.} Skin tissue serves as a valuable application, given contact with external surfaces increases risk for potential bacterial exposure. We study the delivery of recombinant human EGF protein (rhEGF) to a wounded human keratinocyte cell culture in vitro, leveraging nanoparticle-loaded silk sutures as a therapeutic delivery mechanism. Equipped with self-assembly properties, silk fibroin sutures are promising for handling dynamic terrains of a healing epidermal layer during differentiation, enhancing binding to polymeric and nanoparticle solutions, and controlling degradation with strong biocompatibility. We synthesize Dextran Sulfate Sodium - Protamine Sulfate (henceforth referred to as DSS-protamine) nanoparticles with rhEGF to develop a cationic-anionic polyelectrolyte complex. Both compounds have reported antimicrobial activity, which may aid in limiting biofilm growth -- a vital mitigant to bacterial infection risk in skin wounds. Keratinocytes are first prepared in a growth medium and replaced to a basal medium 24h prior to experimentation. The keratinocytes are wounded with a 2-3mm chilled aluminium rod, and the suture is placed on the resulting wound bed. The complete experimental procedure is included in Appendix \ref{lab-procedure}. 

\paragraph{Polymer Discovery with Latent Space Similarity Search and Filtering.}

A SMILES string corresponding to a state-of-the-art polymer topcoat (PLGA) was obtained, and converted to PSELFIES as described above. We applied the trained PSELFIES encoder to the resulting string to obtain a latent space embedding. Nearest neighbors search ($k=100$) was performed to identify similar polymers -- both in terms of PSELFIES and  graph, as a result of cross-modal alignment. The resulting polymer candidates exhibited 92.59-97.52\% similarity to the PLGA representation. 

To filter on the basis of accessibility and feasibility of synthesis, Ertl-Schuffenhauer synthetic accessibility scores \citep{Ertl2009SAscore} were obtained (included in the PI1M dataset), and candidates with a score $\geq 5$ were filtered out. We then calculated the Tanimoto similarities between fingerprints of the remaining polymers. Subsequently, we developed a retrieval-augmented generation (RAG) pipeline to leverage a large language model (LLM) for reasoning about the candidates and clustering them with textual labels for interpretation. Specifically, SMILES (converted to fingerprint), SELFIES, and a textual description were sourced from the ChEBI-20-MM dataset \citep{Liu2025KnowledgeLearningPreferences}, embedded using the OpenAI \verb+text-embedding-3-small+ embedding model \citep{OpenAI2024NewEmbeddingsAPIUpdates}, and stored in a vector index with Weaviate. Given the candidate and the 10 most similar entries in the vector database, the LLM --- specifically, GPT-5 \citep{singh2025openaigpt5card} --is prompted to extract a motif signature from the SELFIES syntax and the textual description. This yields information such as common functional groups, aliphatic vs. aromatic hydrocarbons, etc. For each candidate, we then apply our property prediction model to obtain inferred values for the 8 properties considered. 
Then, given the motifs and property values, the model is asked to propose clusters and assign all candidates to these groups -- this yielded 8 clusters (described in detail in Appendix \ref{rag-desc}). Notably, the majority of the remaining polymers belonged to the following groups: 
\begin{enumerate}
    \item PLGA-like aliphatic polyesters with compact di-ester repeats
    \item Carbonate-containing with –O–C(=O)–O– motif
    \item PCL-containing longer aliphatic spacers
\end{enumerate}

The LA-TMC (group 2) and PLLA-PCL (group 3) polymer families were selected as representative of these groups, along with a PLGA (65:35) blend baseline. The polymers were sourced from a chemical manufacturer, and will be henceforth referred to by their composition. 

\begin{figure}
    \centering
    \includegraphics[width=0.8\linewidth]{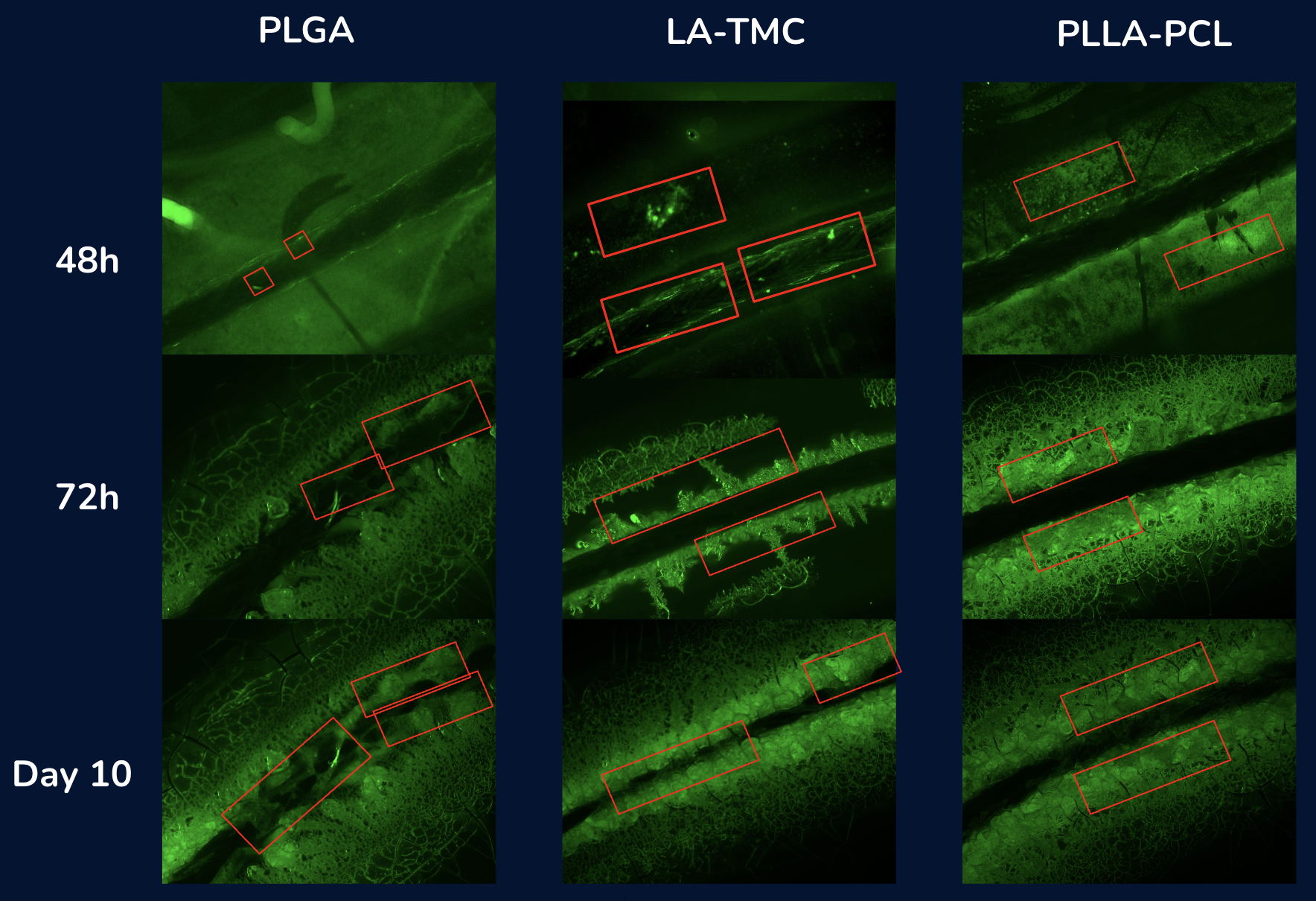}
    \caption{Immunofluorescence imaging of wound-healing assays following rhEGF delivery via polymer-coated, nanoparticle-loaded sutures at 48h, 72h, and 10 days post-experimentation; red boxes indicate crucial points of qualitative analysis toward targeted tissue regeneration. The LA-TMC images demonstrate a wider region of cell activity toward the wound bed, visible keratinocyte presence on the rhEGF coated suture at 48h, and mediated cell migration underneath the suture.}
    \label{fig:if-analysis}
\end{figure}

\paragraph{Immunofluorescence and ELISA Demonstrate Controlled rhEGF Release.}

To assess the efficacy of each suture scaffold, we investigated keratinocyte differentiation with rhEGF delivery and its effect on wound healing, with (1) immunofluorescence analysis for qualitative evidence, and (2) ELISA analysis to study the payload delivery.  
We conducted an immunofluorescence analysis at 0h, 48h, and 72h, with well plates consisting of sutures coated with each of PLGA, LA-TMC, and PLLA-PCL. The keratinocytes were stained utilizing the Ki-67 monoclonal antibody in a 1:200 ratio with a Bovine Serum Albumin (BSA) blocking buffer, marking its expression during mitosis or G$_2$ phase via activation with the MKI67 gene. 

Our results from the immunofluorescence analysis in Figure \ref{fig:if-analysis} suggest that at 48h, 
keratinocyte binding to the suture was primarily and most successfully achieved by the LA-TMC coated suture, followed by PLLA-PCL and PLGA. This conclusion was made based on a qualitative analysis of fluorescing cells, highlighting those attached to the suture (indicating localized exposure to the rhEGF payload) as well as those lined along each epithelial tongue, preparing for cross-bed differentiation. 

At 72h, once again, the goal of \textit{controlled} keratinocyte movement across the wound bed---aided by the rhEGF-carrying suture---is primarily  fulfilled by LA-TMC. Upon a general analysis, we observed  significant keratinocyte growth and migration of cell formations across the wound bed, as well as cell linings on either epithelial tongue actively going through the differentiation process. While PLGA and PLLA-PCL exhibited suboptimal growth patterns with uncontrolled development in certain regions, LA-TMC demonstrated targeted and consistent growth toward the wound bed. 

\begin{figure}
    \centering
    \includegraphics[width=0.8\linewidth]{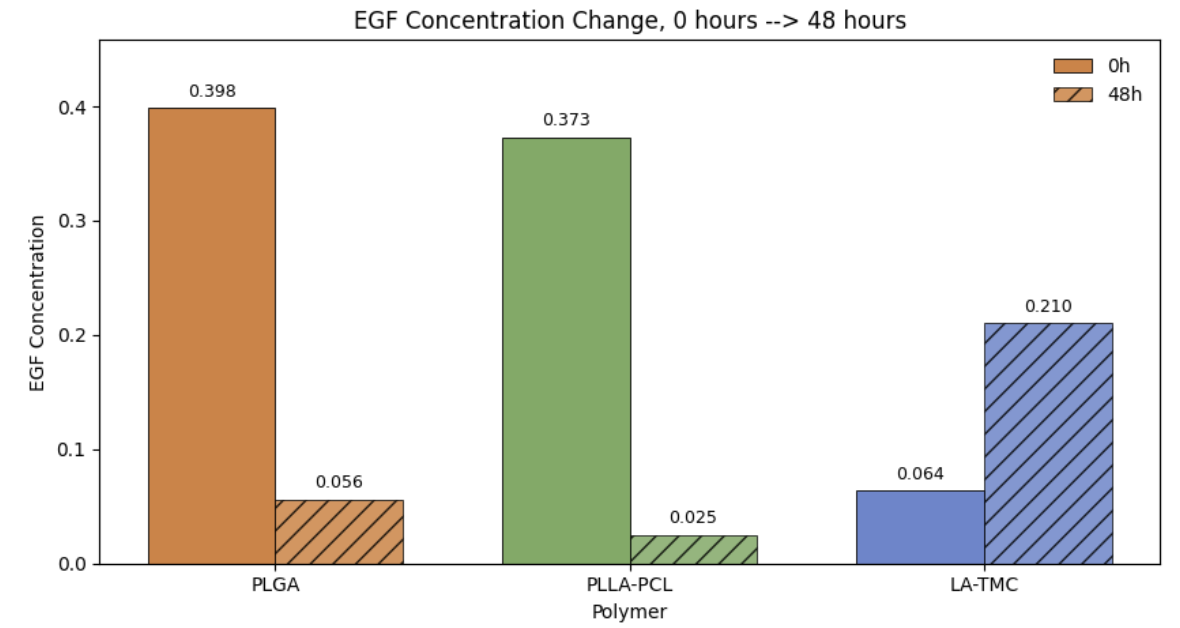}
    \caption{Results of the EGF ELISA analysis for each polymeric suture topcoat from 0h $\rightarrow$ 48h. These findings demonstrate an evident increase in EGF concentration with LA-TMC, suggesting more deliberate and controlled release mechanics, and a clear decrease with PLGA and PLLA-PCL.}
    \label{fig:elisa-egf}
\end{figure}

To quantify rhEGF release into the supernatant, we utilized a commercially acquired EGF ELISA kit, specialized to immobilizing supernatant antibodies to quantify the EGF concentration. The 0h and 48h 250µL supernatant samples were flash-frozen overnight at -80\textdegree C before distributing 50µL into each of 12 wells (including duplicates). Along with the ELISA listed calibration standards, the plate was read at a 450nm absorbance level, allowing us to use the reported absorbance for each standard and develop an EGF concentration calibration curve  (Figure \ref{fig:elisa-calibration}). The resulting quadratic regression equation was then used to correlate the respective ELISA absorbances for the polymer-specific supernatant samples (at 0h and 48h) and calculate approximate EGF concentrations. Each pair of results is plotted in Figure \ref{fig:elisa-egf}, demonstrating the gradients of supernatant EGF concentration. The LA-TMC sample contained little rhEGF in the supernatant at 0h, but the concentration exhibited an +330\% increase by 48h, compared to losses of -86\% and -93\%, in PLGA and PLLA-PCL. respectively. Overall, from an rhEGF delivery standpoint, LA-TMC was the sole polymer coating that facilitated controlled and targeted release, as a sustained release rate is optimal for therapeutic payload delivery in wound healing. Accelerated healing and wound closure can motivate future testing of earlier removal times in vivo while  minimizing bacterial exposure.   

 In Appendix \ref{appendix:atr-ftir}, we include ATR-FTIR spectroscopy analysis on the nanoparticle complex and for polymer-EGF interactions; we find compelling evidence for electrostatic bonding within the nanoparticle complex, motivating its role as a vessel to facilitate polymer-EGF ligand dimerization. 

\section{Discussion}

We introduce GenPoly, a framework that facilitates polymer discovery, including multimodal pre-training and polymer property prediction, along with candidate generation, filtering, and clustering. This yields bioresorbable polymers that can be leveraged for biomedical applications, as demonstrated through in vitro wound healing experimentation. 
We highlight the generality of GenPoly: it can be used to generate (possibly novel) polymers for other applications, including as scaffolds for complex, in situ tissue regeneration settings. 
Advancements in data-efficient property prediction, as well as data access for a wider range of properties (e.g. solubility, tensile strength, etc.) could be valuable to refine candidate selection, with clearer intuition of the strengths and weaknesses of each polymer.
While we leverage property prediction as part of the candidate filtering process, property-aware guidance in decoding (e.g. via latent diffusion) could present further improvements. Simultaneously, we envision that improvements to our decoding strategy, which uses a state-of-the-art polymer as an anchor, as well as access to larger polymer corpora (or using synthetic data generation) could discover new spaces of underexplored, yet viable and synthesizable polymers. 

\bibliography{iclr2026_conference}
\bibliographystyle{iclr2026_conference}
\newpage
\appendix

\section{In Vitro Experimental Procedure}\label{lab-procedure}

For the experimentation on epithelial tissue, the 500µL Primary Epidermal Keratinocytes(Normal, Human, Adult (HEKa)) sample was sourced from the American Type Culture Collection (ATCC). All wet lab experimentation was performed in a Biosafety Level 1 (BSL-1) lab in accordance with the CDC/NIH biosafety levels.

\paragraph{Keratinocyte Culture Expansion and Seeding in Growth Medium.}

A 300 mL keratinocyte basal medium was prepared using Keratinocyte Serum-Free Medium (K-SFM; Gibco, Thermo Fisher Scientific) supplemented with 3 mL bovine pituitary extract (BPE; Gibco) and 0.6 mL recombinant human EGF (rhEGF; Gibco), then pre-warmed to 37\textdegree C. Medium (5 mL) was added to a 
T25 flask (Thermo Fisher Scientific) and equilibrated overnight in a humidified incubator (Quincy Lab model 12-140) at 37\textdegree C. Frozen keratinocytes were rapidly thawed at 37\textdegree C, and 1 mL of the cell suspension was transferred into 5 mL pre-warmed complete K-SFM in the T25 flask. Additional medium was added to yield a final volume of 10 mL, and the cells were incubated overnight at 37\textdegree C. For seeding, culture medium was aspirated and cells were washed once with Dulbecco’s phosphate-buffered saline (DPBS; Gibco). Cells were detached using 1 mL Trypsin-EDTA (Thermo Fisher Scientific) and neutralized with an equal volume of Trypsin Neutralizing Solution (ATCC) to form a single-cell suspension. The suspension was transferred to a 15 mL tube and brought to 12 mL with pre-warmed medium. Cells were seeded at 1 mL per well (target $\approx 2.5 \times 10^4$ cells per well) in a multi-well plate and incubated at 37\textdegree C for 6 hours to allow for attachment. To eliminate exogenous EGF and ensure that solely suture-delivered EGF is quantified, the medium was replaced after 6 hours with an rhEGF-free assay medium consisting of K-SFM supplemented with only BPE, followed by overnight incubation at 37\textdegree C. 

\paragraph{Synthesis of DSS-protamine Nanoparticle Solution.}

Dextran sulfate sodium (DSS; Fisher Scientific) and Protamine (Fisher Scientific) were prepared in DPBS at 0.36 mg mL${}^{-1}$ and 0.30 mg mL${}^{-1}$, respectively, corresponding to a 1.2:1 (DSS:protamine) mass ratio. Specifically, DSS (0.36 mg) was dissolved in 10 mL DPBS, and protamine (0.30 mg) was dissolved in 10 mL DPBS. Subsequently, 20 µg of lyophilized recombinant EGF was then reconstituted in 1 mL DPBS and combined with the protamine solution. The DSS solution was then added dropwise to the protamine–EGF mixture in a sterile 15 conical mL tube and gently mixed. The mixture was incubated at room temperature for 10 minutes to allow adequate formation of the nanoparticle / polyelectrolyte complex, then maintained on wet ice until use for silk fibroin suture coating.

\paragraph{Dissolution of Polymeric Topcoats and Suture Coating. }

Sterile silk fibroin sutures (Perma-Hand) were cut to 3.5 cm segments (to match the well diameter), then placed on sterile Parafilm until coating.
Under a fume hood, PLGA (Sigma-Aldrich), LA–TMC (EVONIK LT 706S), and PLLA–PCL (EVONIK LC 703S) were each dissolved in hexafluoroisopropanol (HFIP; Sigma-Aldrich) to 10 mg mL${}^{-1}$ and dispensed into separate metal weighing boats. Using forceps disinfected with 95\% ethanol, suture segments (3 per polymer) were immersed in the corresponding polymer solution for 2 min and subsequently air-dried. This dip-coating and drying cycle was repeated three times, after which sutures were dried at room temperature for 1 hour on fresh sterile Parafilm.
For nanoparticle loading, the nanoparticle suspension was aliquoted into three sterile boats, corresponding to each polymer group. Polymer-coated sutures were immersed in the corresponding nanoparticle suspension and maintained on wet ice for 20 min with gentle agitation to facilitate uniform coating. Sutures were then removed, allowed to drain briefly to remove excess suspension, and then transferred to fresh sterile dishes until application to cells.

\paragraph{Cell Monolayer Wounding Simulation and Coated Suture Placement.}

For wound-healing assays, an aluminium foil rod (2-3 mm diameter) was pre-chilled at -80\textdegree C and DPBS and rhEGF-free assay medium were pre-warmed to 37\textdegree C. Culture plates were removed from the incubator, the existing medium was aspirated, and a scratch wound was introduced in the center of each well by briefly contacting the confluent monolayer with the pre-chilled aluminium rod for 3-5 seconds. Immediately after the introduction of the wound, wells were washed with 1 mL pre-warmed DPBS to remove detached cells. DPBS was subsequently aspirated, and 1 mL pre-warmed rhEGF-free assay medium was added. Using forceps disinfected with 95\% ethanol, one coated suture was placed per well across the wound region, such that the suture lay flat and spanned the wound end-to-end. Plates were returned to a humidified incubator at  37\textdegree C and sampled at 20 minutes (defined as 0 hours), 48 hours, and 72 hours post-placement. 

\subsection{EGF ELISA Calibration}

\begin{figure}[h]
    \centering
    \includegraphics[width=\linewidth]{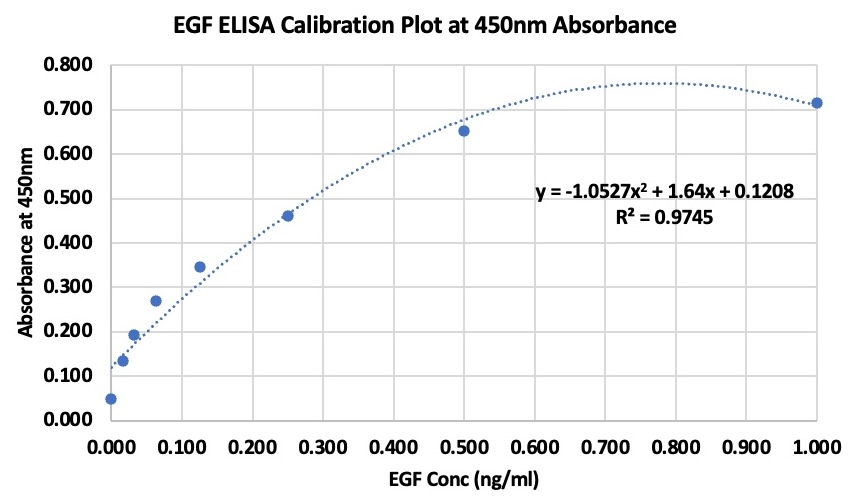}
    \caption{ELISA calibration curve for rhEGF measured at 450nm absorbance, then used to quantify EGF concentrations in polymer-conditioned samples.}
    \label{fig:elisa-calibration}
\end{figure}

To support the quantification of EGF in polymer-conditioned supernatant samples, we first generated a standard calibration curve using the commercial kit-provided ELISA standards upon reading the plate at 450 nm. A quadratic regression was fit to the standard absorbance values and used to convert the sample absorbances of 0h and 48h samples into corresponding EGF concentrations. The resulting calibration curve and regression are shown in Figure \ref{fig:elisa-calibration}. 

\subsection{ATR-FTIR Spectroscopy Analysis of Nanoparticle Complex}\label{appendix:atr-ftir}

To more effectively ascertain the specific chemical properties that enhanced the polymer-nanoparticle binding at differing levels in accelerated and controlled rhEGF delivery, we utilized Attenuated Total Reflectance-Fourier Transform Infrared (ATR-FTIR) spectroscopy to uncover the dynamic changes in intramolecular and electrostatic forces that occurred in solution.
Each polymer solution was re-created in a 10 mg /mL ratio, and likewise the DSS-protamine nanoparticle solution was re-created in a 1.2:1 ratio without the inclusion of the rhEGF protein. 

\paragraph{Spectral Shifts in DSS-protamine Complex Indicate Potential N-H Bond Strengthening due to Electrostatic Interactions. }
To analyze the composition of DSS-protamine nanoparticles, we first examined dissolved DSS, dissolved protamine (both in D-PBS), and the combination forming the 1.2:1 nanoparticle solution. The resulting stack from the first-pass FTIR analysis demonstrated 3 primary spectral changes in the nanoparticle solution, relative to the component DSS and protamine that indicate potential chemical interactions leading to strong binds between the two. Moreover, a stronger DSS-protamine bond develops the notion that a cationic rhEGF binds to a complex with a strong ionic bonding capacity, leading to its electrostatic retainment within the nanoparticle matrix. 
These spectral shifts are consistent with ion-pairing within the DSS–protamine complex, which may contribute to controlled rhEGF release through a mechanism such as ionic exchange to the EGFR receptor in keratinocytes, rather than immediate burst release. 

\begin{figure}[H]
  \centering

  \begin{subfigure}[t]{0.49\textwidth}
    \centering
    \includegraphics[width=1.1\linewidth]{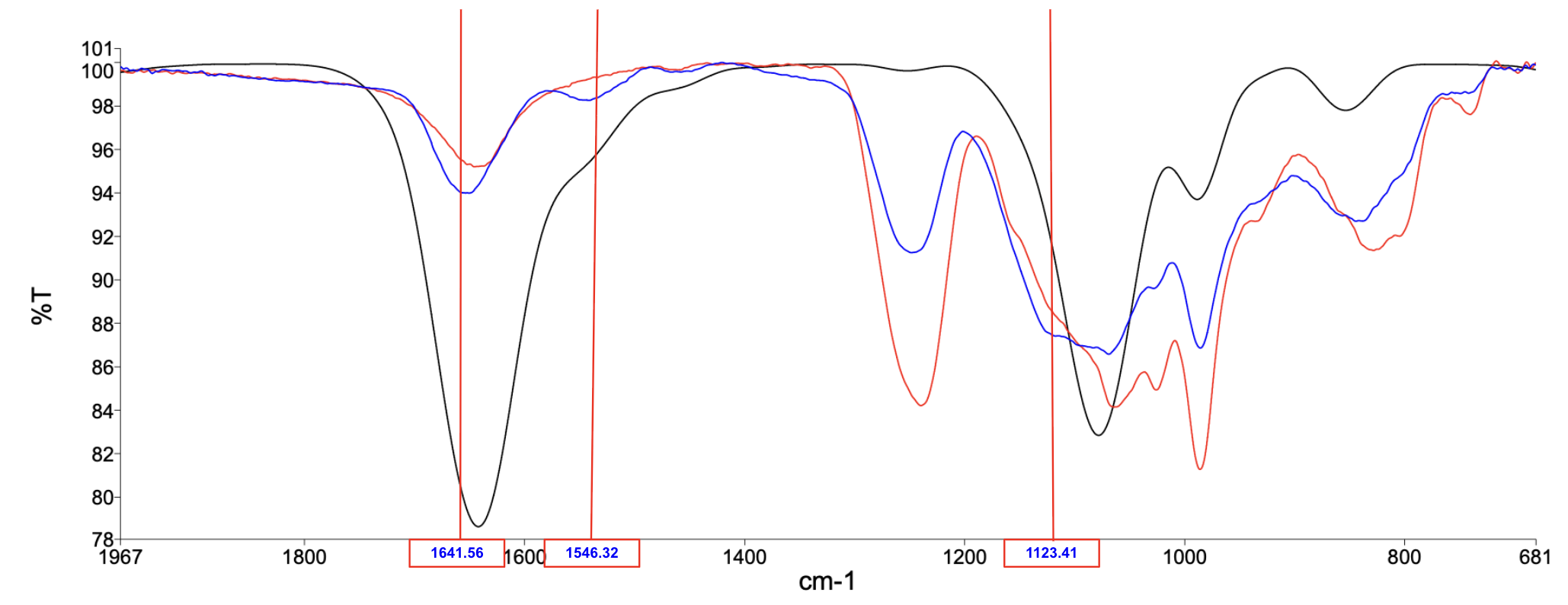}
    \caption{}
    \label{fig:np-broad}
  \end{subfigure}
  \hfill
  \begin{subfigure}[t]{0.49\textwidth}
    \centering
    \includegraphics[width=0.9\linewidth]{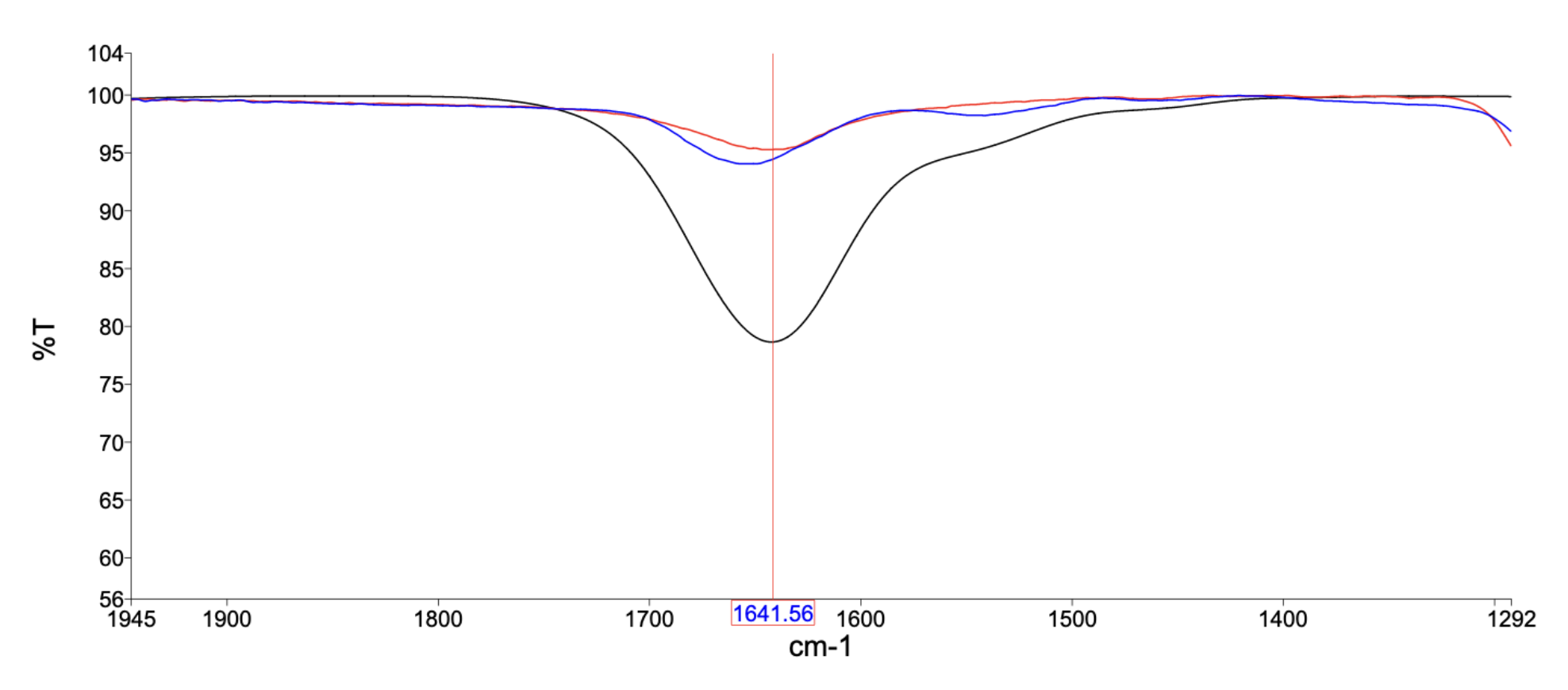}
    \caption{}
    \label{fig:np-zoom}
  \end{subfigure}

  \caption{(a) Stacked FTIR spectra of the 1.2:1 DSS-protamine nanoparticle complex (blue), along with reference standards of protamine in D-PBS (black) and DSS in D-PBS (red); (b) Expanded view of N-H stretching region, with band shifts associated to ion-pairing between DSS and protamine subgroups. }
  \label{fig:dss-prot-ftir}
\end{figure}

From the overlaid analysis in Figure \ref{fig:np-broad}, the three potential spectral bands at regions of observed wavenumber position shifts, namely at 1641.56 cm${}^{-1}$, 1546.32 cm${}^{-1}$, and 1123.41 cm${}^{-1}$, correspond to shifts of the N-H bend (C-N stretch) of the amide II band, C=O amide I band, and C-O amide II band \citep{niu_ir_frequencies}. Ultimately, the N-H band, reflected in Figure \ref{fig:np-zoom}, was chosen between the ~1600 cm${}^{-1}$ NH bond region and C=O bond region, primarily due to the chemical composition of protamine that properly reflects the guanidinium and N-H modes that facilitate polyelectrolyte complexation with a non-N-H-group-containing DSS. Therefore, the observed band shift in the nanoparticle complex spectrum, relative to the properly aligned protamine and DSS spectra, reflects ion-pairing and electrostatic interactions between the DSS sulfate groups and protamine's arginine guanidinium chains \citep{Vazdar2018ArginineMagic}, which stretch the complex's N-H bonds, thereby increasing its strength and reducing free H-bonding, forming an anionic electrostatic microenvironment for binding to a cationic rhEGF. 

\begin{figure}[h]
    \centering
    \includegraphics[width=\linewidth]{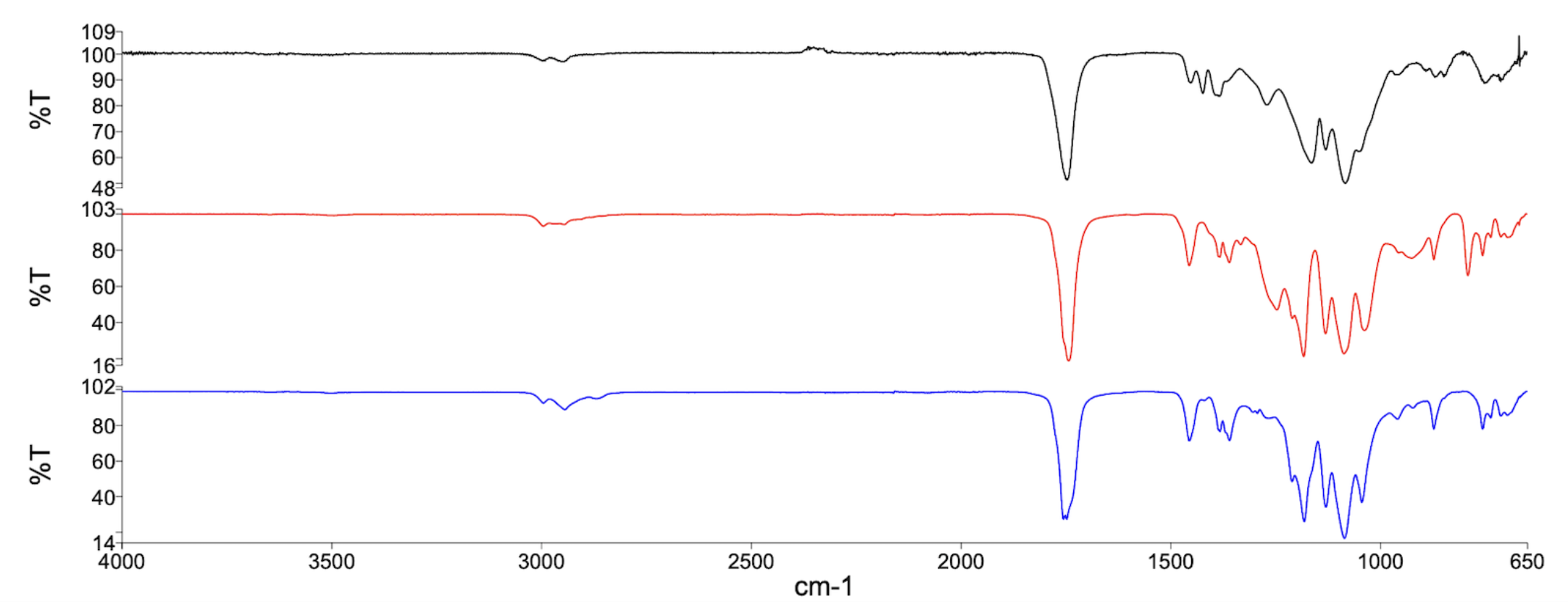}
    \caption{Stacked ATR-FTIR spectra of PLGA (black), LA–TMC (red), and PLLA–PCL solid polymers (blue).}
    \label{fig:ftir-polymer-solids}
\end{figure}

\paragraph{Ligand-Induced Dimerization of rhEGF by the Nanoparticle-Polymer Matrix }

Upon comparison of each polymer's chemical functional groups, we evaluated the interaction mechanisms specific to those that are know to electrostatically complex with protein ligands, particularly in certain regions of the stacked spectra of solid polymer samples in Figure \ref{fig:ftir-polymer-solids}.

A strong C=O stretching band was observed across all polymers in the 1740–1760 cm${}^{-1}$ region, which accurately corresponds to their individual chemical structures. However, the carbonyl band intensity of  LA-TMC uniquely demonstrated subtle peak broadening relative to PLGA and PLLA-PCL, indicating a more polar environment. This is particularly relevant for rhEGF binding, as rhEGF contains charged and hydrogen-bonding residues that can interact favorably with carbonyl-rich polymeric chains \citep{Schlessinger2002LigandInducedEGFR}.

In addition, LA-TMC demonstrated pronounced alkyl C–H stretching modes ($\approx2850–3000$ cm${}^{-1}$) alongside a denser fingerprint region ($\approx1000–1300$ cm${}^{-1}$), reflecting a balanced combination of hydrophobic chain segments \citep{raman-spectra}. Regarding protein delivery, hydrophobic domains promote polymer–protein association, and simultaneously contain polar ester groups that enable non-covalent interactions that avoid denaturation \citep{LYU2023123110}. By contrast, PLGA displayed a relatively simpler spectral profile with fewer potential interaction-indicating regions, while PLLA-PCL exhibited a stronger hydrophobic nature but reduced apparent polarity, which suggests a more limited ability to maintain stable rhEGF association.

Overall, the chemical compositions revealed by the FTIR analysis ultimately conclude that LA-TMC may provide a more interaction-permissive environment for rhEGF binding through enhanced hydrogen bonding and electrostatic interactions, while preserving sufficient hydrophobic character for ligand retention. We posit that this interaction underlies the promotion of effective rhEGF to EGFR signaling, possibly by facilitating receptor dimerization and sustained activation of receptor tyrosine kinases (RTKs) \citep{Petrova2025EGFRKinaseDimerization}, though this is beyond the scope of this work. 

\section{Model Architecture}\label{architecture}

\subsection{Multimodal Learning (Polymer Strings, Molecular Graphs)}

\begin{figure}[H]
    \centering
    \includegraphics[width=\linewidth]{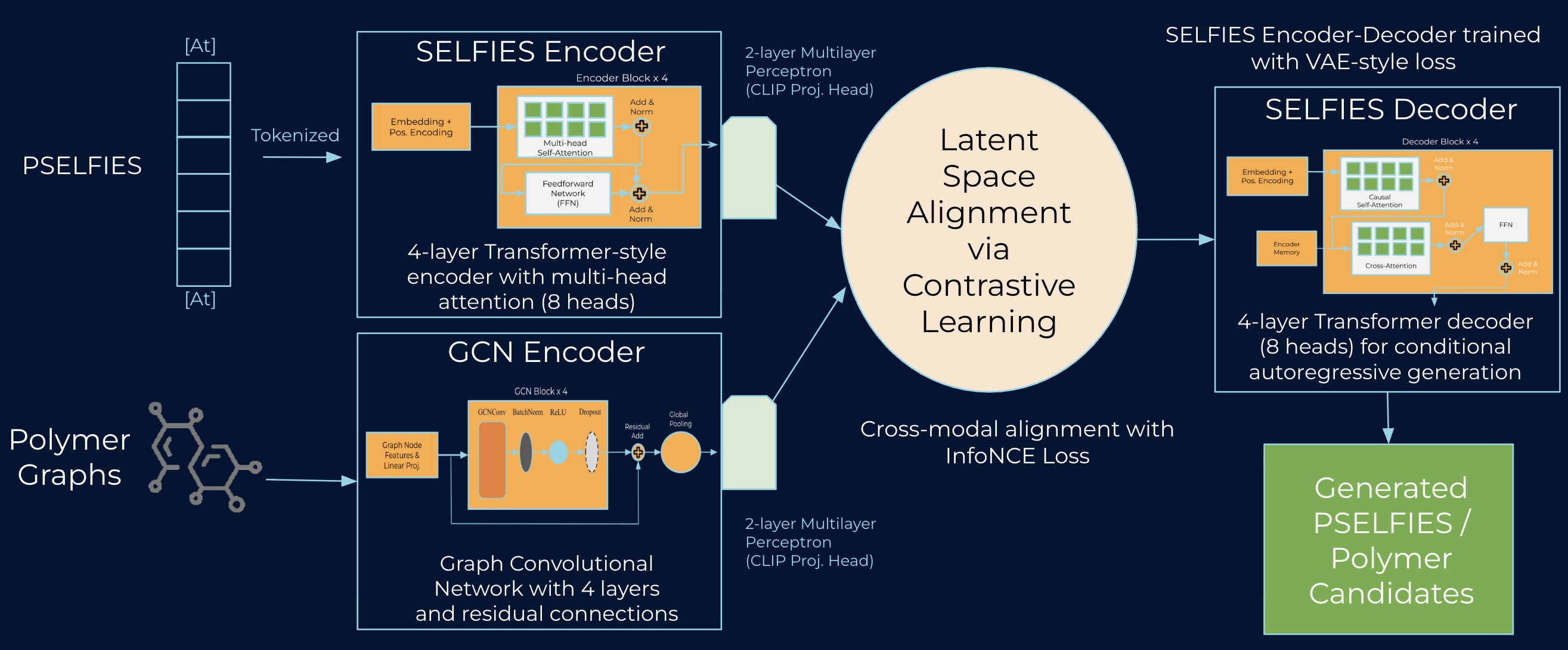}
    \caption{Visualization of the end-to-end multimodal pre-training pipeline. }
    \label{fig:architecture}
\end{figure}

\begin{figure}[H]
  \centering

  \begin{subfigure}[t]{0.49\textwidth}
    \centering
    \includegraphics[width=0.95\linewidth]{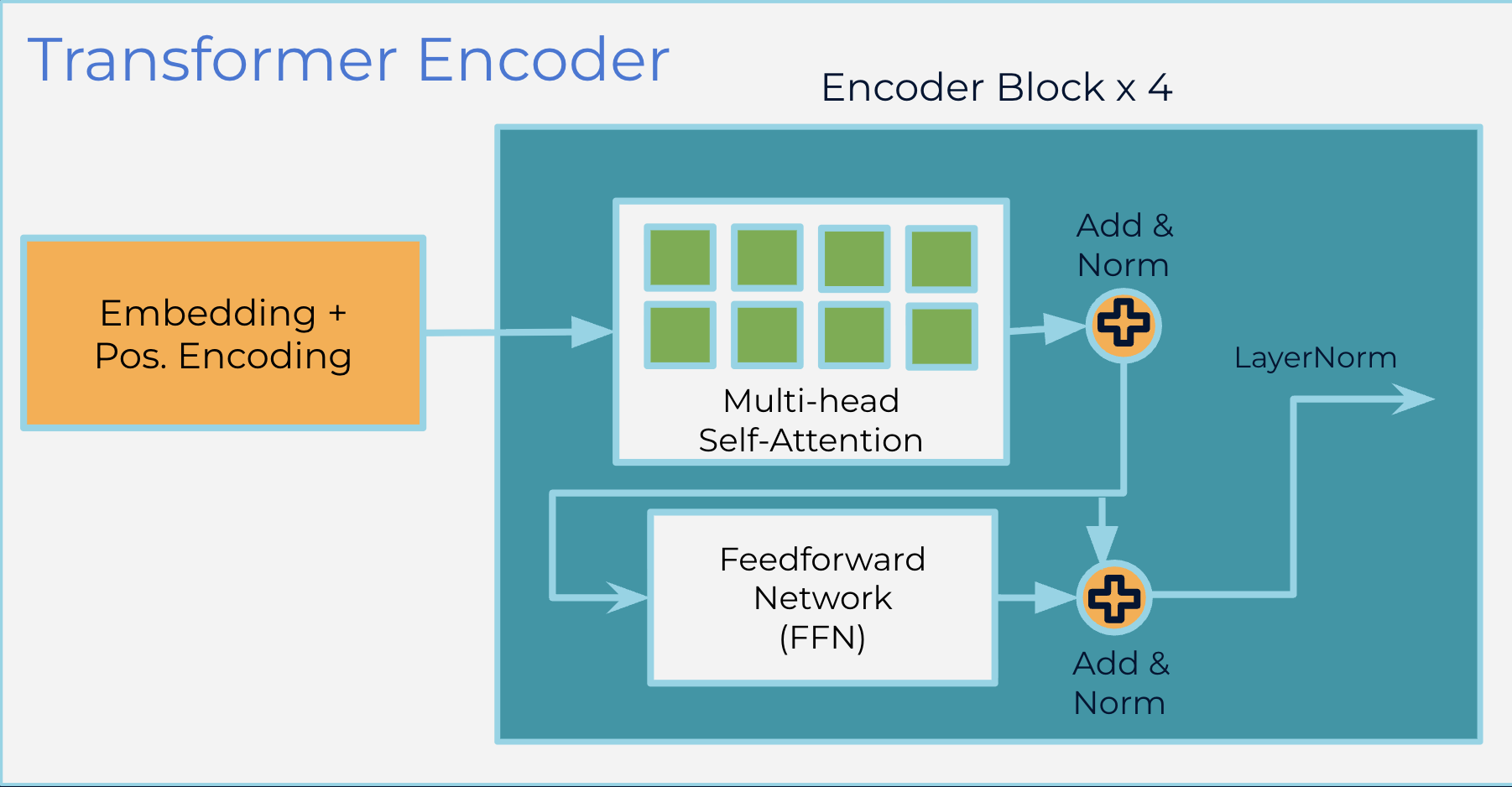}
    \caption{}
    \label{fig:enc}
  \end{subfigure}
  \hfill
  \begin{subfigure}[t]{0.49\textwidth}
    \centering
    \includegraphics[width=0.95\linewidth]{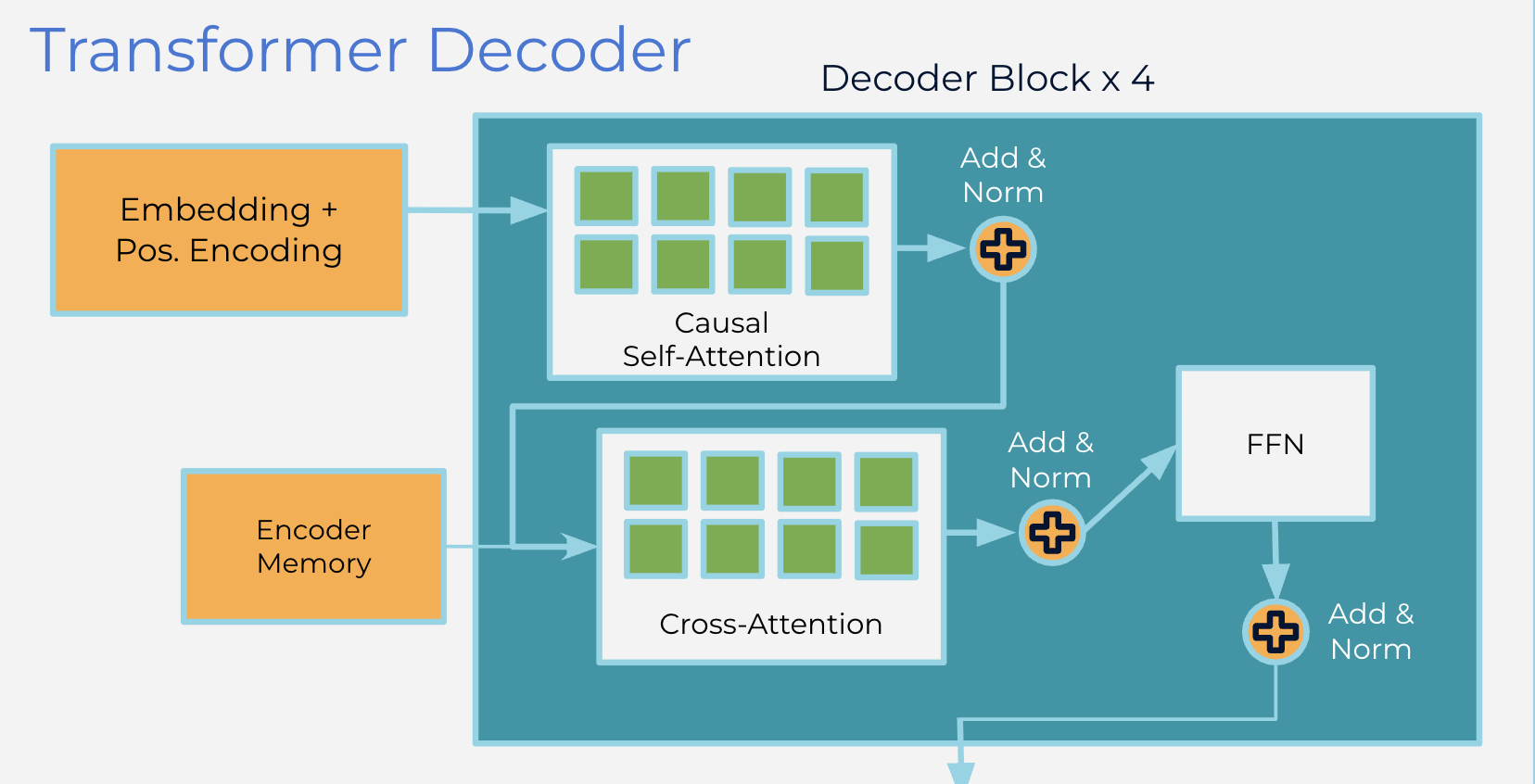}
    \caption{}
    \label{fig:dec}
  \end{subfigure}

  \vspace{0.5em}

  \begin{subfigure}[t]{\textwidth}
    \centering
    \includegraphics[width=0.6\linewidth]{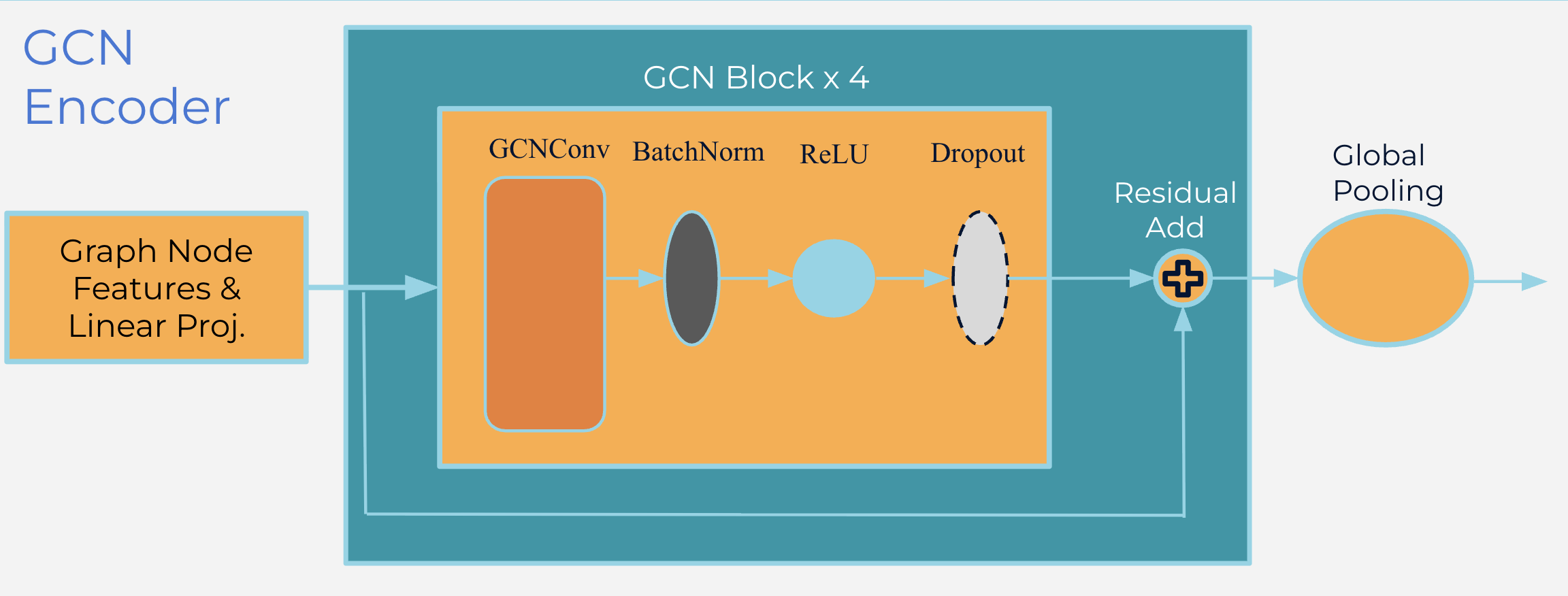}
    \caption{}
    \label{fig:gcn}
  \end{subfigure}

  \caption{Visualizations of Transformer Encoder architecture for SELFIES (a), Transformer Decoder architecture for autoregressive SELFIES generation (b), and GCN Encoder for graphs (c). }
  \label{fig:encs-and-dec}
\end{figure}

As described in Section \ref{main:method}, our multimodal foundation model is pre-trained over polymer SELFIES (PSELFIES) and graph representations. The complete pipeline is depicted in Figure \ref{fig:architecture}, with the encoders and decoder shown in greater detail in Figure \ref{fig:encs-and-dec}. The polymer graph is encoded by a GCN that first projects node features to a hidden space and then applies $4$ residual GCNConv blocks with batch normalization and dropout. 
Graph-level features are obtained by concatenating global mean-pooled and max-pooled node states. Simultaneously, given a tokenized PSELFIES string, we train a 4-layer Transformer encoder to  produce contextualized token states which are masked and globally pooled. We obtain sequence embeddings from each in a shared embedding space by sampling from a Gaussian $q_{\phi}(z_i | i)$ for $i \in \{s,g\}$, corresponding to (P)SELFIES and graphs, respectively. KL penalties were applied to each Gaussian for latent space regularization, and then each sample is passed this through an multilayer perceptron (MLP) projection head. 

Cross-modal alignment is learned with an InfoNCE loss (discussed in Section \ref{main:method},) computing all-pairs similarities over a matrix $\ell$ and applying bidirectional cross-entropy (PSELFIES $\xleftrightarrow{}$ Graph). In training the 4-layer Transformer decoder\footnote{Note while the SELFIES decoder is indeed trained for autoregressive generation of SELFIES, we primarily use nearest neighbors search in this work for candidate generation. As a result, the encoder-decoder formulation serves as a backtranslation mechanism to  refine the SELFIES encoder for improved latent space representation.} with a VAE-style reconstruction loss, we leverage cross-attention to a token $z$ obtained from a linear projection of $z_s$. The final trained model is 7.84M parameters using a single NVIDIA A100 GPU accessed via Google Colaboratory Pro. 

\subsection{Property Prediction}\label{property-prediction}

We perform property prediction using 8 property values, with data sourced from the Khazana Polymer Informatics dataset, and adopt a similar methodology as \citep{KUENNETH2021100238}. The complete list of properties is: (1) Atomization Energy (Eat), measured in Electron Volts per Atom (eV/atom); (2) Crystallization Tendency (Xc) measured in percentage; (3-6) Band Gap Chain (Egc), Band Gap Bulk (Egb), Electron Affinity (Eea), and Ionization Energy (Ei), measured in Electron Volts; (7-8) Refractive Index (nc) and Dielectric Constant (eps), which have a reference value of 1. The property prediction model is applied on the shortlist of candidates generated from the decoder via nearest neighbors search, to inform the filtering and clustering process.

The learning objective is a multi-task mean-squared error loss, representing the sum over the task-specific MSE losses:
\[
\mathcal{L} = \sum_{t=1}^T \frac{1}{|\mathcal{D}_t|} \sum_{i \in \mathcal{D}_t} (\hat{y}_{i,t} - y_{i,t})^2
\]

\begin{figure}[H]
    \centering
    \includegraphics[width=0.5\linewidth]{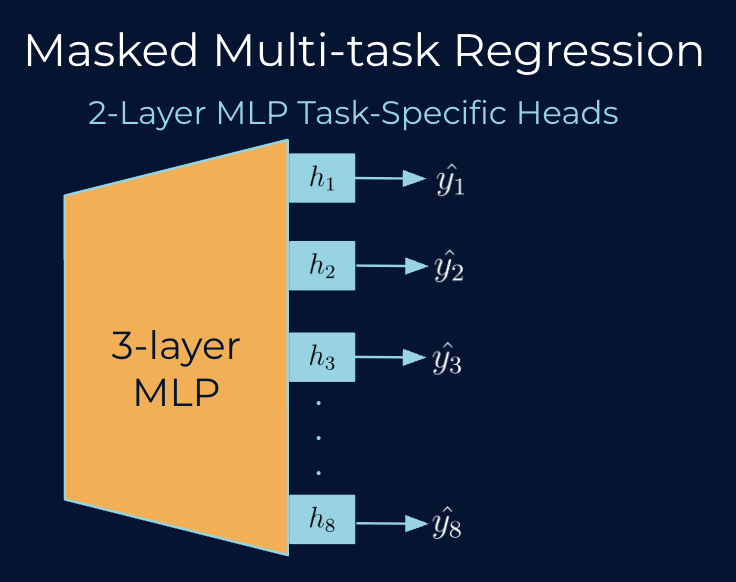}
    \caption{Visualization of the property prediction architecture, consisting of a shared MLP along with task-specific MLP heads. }
    \label{fig:placeholder}
\end{figure}

\section{Training Curves and  Hyperparameters}\label{curves-hyperparameters}

\begin{figure}[h]
  \centering

  \begin{subfigure}[t]{0.49\textwidth}
    \centering
    \includegraphics[width=0.8\linewidth]{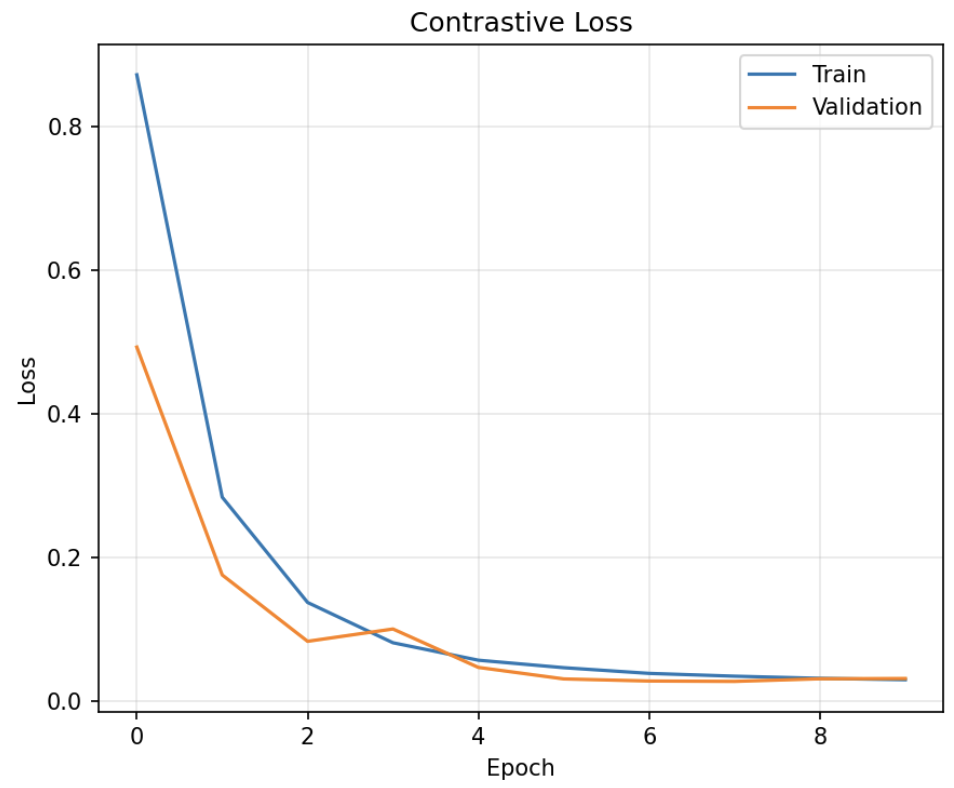}
    \caption{}
    \label{fig:contrastive}
  \end{subfigure}
  \hfill
  \begin{subfigure}[t]{0.49\textwidth}
    \centering
    \includegraphics[width=0.8\linewidth]{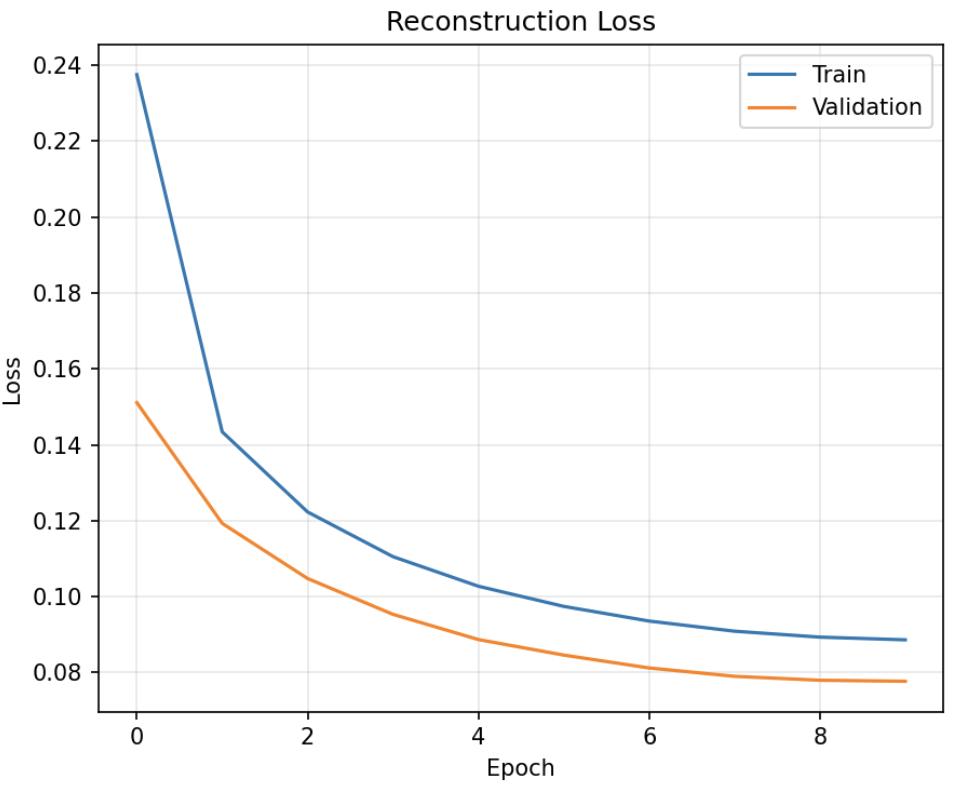}
    \caption{}
    \label{fig:reconstruction}
  \end{subfigure}

  \vspace{0.5em}

  \begin{subfigure}[t]{\textwidth}
    \centering
    \includegraphics[width=0.5\linewidth]{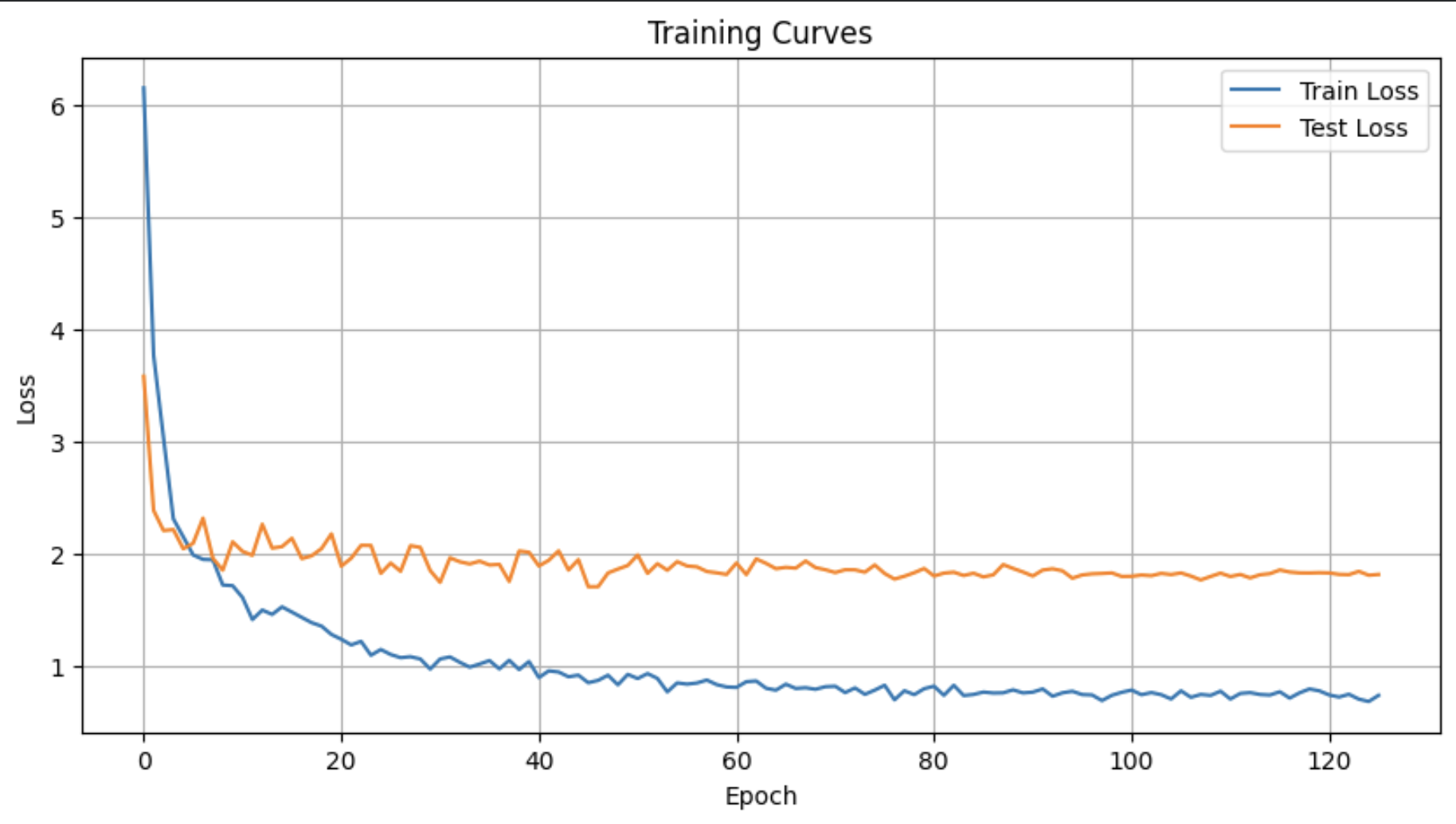}
    \caption{}
    \label{fig:prop-pred}
  \end{subfigure}

  \caption{(a) Plot of contrastive (CLIP-style) loss, (b) plot of reconstruction (VAE-style) loss over 10 epochs of training our multimodal model on $\approx 800,000$ polymers, along with validation loss. (c) Plot of multi-task mean-squared error loss in polymer property prediction, with test loss. }
  \label{fig:loss-curves}
\end{figure}

We include the training curves in the GenPoly pipeline below, for both the multimodal model (Figures \ref{fig:contrastive} and \ref{fig:reconstruction}) and property prediction (Figure \ref{fig:prop-pred}). We use an 80/20 train-validation split for pre-training over polymers, and a 90/10 train-test split for property prediction, given limited data.  

The hyperparameters used in training are included in Table \ref{tab:final_mm_hyperparams} (multimodal pre-training) and Table \ref{tab:final_prop_hyperparams} (property prediction).  

\begin{table}[H]
\centering
\caption{Final hyperparameters for the pre-trained multimodal model.}
\label{tab:final_mm_hyperparams}
\begin{tabular}{|l|c|}
\hline
\textbf{Parameter} & \textbf{Value} \\
\hline
Latent Dim & 256 \\
Hidden Dim & 512 \\
Epochs & 10 \\
Learning Rate & 1e-5 \\
Dropout Parameter & 0.1 \\
Batch Size & 32 \\
Weight Decay & 1e-5 \\
Optimizer & AdamW \\
Graph features & Node features $= 8$, Edge features $= 6$ \\
\hline
\end{tabular}
\end{table}

\begin{table}[H]
\centering
\caption{Final hyperparameters for the property prediction model.}
\label{tab:final_prop_hyperparams}
\begin{tabular}{|l|c|}
\hline
\textbf{Parameter} & \textbf{Value} \\
\hline
Hidden Dims (3-layer MLP) & (512, 256, 128) \\
Hidden Dims (2-layer Heads) & (64, 64) \\
Epochs & 200 (Early Stopping; stopped at 126) \\
Learning Rate & 1e-3 \\
Dropout Parameter & 0.3 \\
Batch Size & 64 \\
Weight Decay & 1e-5 \\
Optimizer & AdamW \\
\hline
\end{tabular}
\end{table}

\section{Analysis of Polymer Clusters}\label{rag-desc}

We include a complete categorization of the polymer candidates following filtering and clustering enabled by the LLM RAG pipeline. Nine candidates were filtered out on the basis of synthetic accessibility, and the remaining 91 polymers were clustered as follows:

\begin{enumerate}
    \item PLGA-like aliphatic polyesters with compact di-ester repeats (28)
    \item Carbonate-containing with –O–C(=O)–O– motif (16)
    \item PCL-containing soft polyesters with longer –CH$_2$– spacers (15)
    \item Nitrogen-containing polyesters (amine/amide/quaternary) (12)
    \item Multi-star, mono-functional, or branch-prone fragments (11)
    \item Unsaturated backbones (vinyl/alkyne) (4)
    \item Silicon-containing fragments (silyl/Si-O) (3)
    \item Phosphoesters with P–O backbones (2)
\end{enumerate}

\begin{figure}[h]
    \centering
    \includegraphics[width=0.95\linewidth]{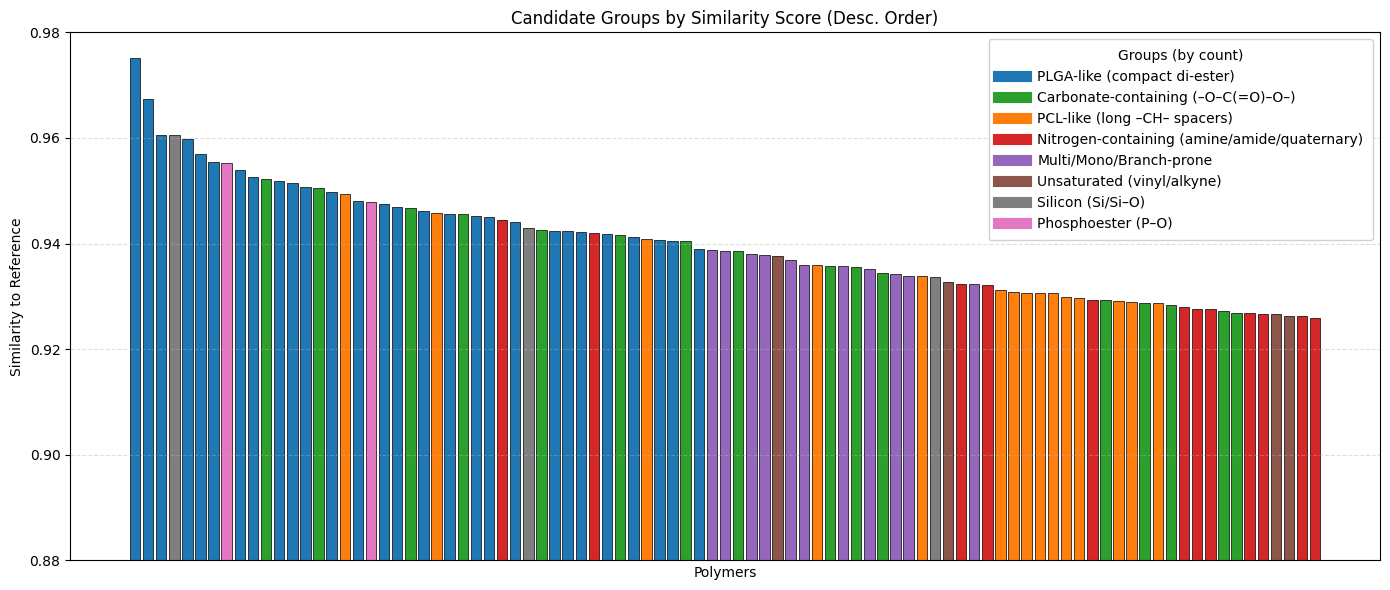}
    \caption{Ranked candidates based on cosine similarity to the reference anchor (PLGA), based on their groupings (color-coded as per the legend). }
    \label{fig:candidates-colored}
\end{figure}

The groups were manually examined and the authors of this work made only two corrections to the groupings (PCL-containing $\rightarrow$ PLGA-like, PLGA-like $\rightarrow$ Multi-star/mono-functional/branch-prone), an inter-annotator agreement with the language model of 97.8\%. 
Polymers for wet lab analysis were chosen from the three groups with the most candidates: "PLGA-like," "Carbonate-containing," and "PCL-like." The polymers belonging to the "Nitrogen-containing polyesters (amine/amide/quaternary)" cluster were deemed to not be appropriate for the in vitro wound healing analysis on keratinocytes given concerns about cytotoxicity. The "multi-star/mono-functional/branch-prone" group's candidates routinely exhibited high crystallization tendency in property prediction, indicating brittleness, while others indicated low ionization energy, suggesting a reactive reduction-oxidation profile which can lead to instability when paired with the nanoparticle complex. The Silicon and Phosphoester clusters, despite having candidates in the top 10 in similarity, have challenges with hydrophobicity and hydrophilicity, respectively: the silicon-containing fragments had a noticeably low predicted dielectric constant while the phosphoesters featured a relatively high predicted dielectric constant, although no noticeable differences in refractive index were predicted. Notably, the former has weak absorbability properties, while the latter can degrade too fast, making them poor fits for our target application.  

\section{Related Work}

\subsection{AI-Driven Polymer Discovery and Property Prediction}

Polymer informatics has benefited from recent advancements in polymer-aware molecular representations that enable large-scale machine learning. While SMILES / PSMILES \citep{Weininger1988SMILES} are foundational for small-molecule cheminformatics, BigSMILES \citep{lin2019bigsmiles} extends this for macromolecular ensembles, and robust string encodings like SELFIES \citep{krenn2020selfies} help to prevent invalid generations. Data availability remains a challenge \citep{audus2017polymerinformatics}, motivating synthetic benchmarks such as PI1M \citep{Ma2020PI1M} and architecture-aware representations that better match polymer structure and property relationships \citep{aldeghi2022ensemblegraph}. \\ \\
Leveraging this data, polymer ML has increasingly adopted pre-training and transfer-learning to enable discovery across tasks and domains. Works such as polyBERT \citep{kuenneth2023polybert} show that polymer-tailored language modeling can yield strong, transferable embeddings for property prediction, while TransPolymer \citep{xu2023transpolymer} demonstrates that Transformer pre-training (i.e. masked language modeling) can improve performance on downstream polymer property benchmarks. Related works such as PolyNC have also studied blending chemical strings with natural language task prompts and multi-task objectives \citep{qiu2024polync}, while self-supervised contrastive learning has been used to stabilize polymer representations and improve robustness \citep{zhou2025polycl}. There also exists works that emphasize goal-directed generation and validation, such as with RL-based multi-objective optimization frameworks \citep{li2026polyrL}, while GPT-style and diffusion-style generation \citep{yang2024polymer_electrolytes} has been applied for de novo polymer electrolyte design. Several efforts pair AI-guided discovery with experimental validation in high-impact applications: optimizing for thermal conductivity \citep{wu2019thermalconductivity}, gas-separation membranes \citep{yang2022gas_separation}, and more recently, for interfacial function to match target adhesive free energies \citep{xu2025adhesive_cgan}.

Polymer property prediction is a core enabler in discovery pipelines, providing a fast screening layer prior to synthesis and wet-lab validation \citep{gao2024mlassisted_design, yue2025mlassisted_molecular_design}. Polymer Genome \citep{kim2018polymergenome} develops a platform prioritizing rapid and accessible property prediction, emphasizing data curation and polymer-aware fingerprints. More recent work has found benefits in shared learning of representations, from sample-efficient multi-task formulations \citep{KUENNETH2021100238} to multimodal representations with textual captions \citep{huang2025unified_multimodal_polymer}. These trends are a key motivator for combining polymer representation learning with lightweight property-specific heads as part of the discovery pipeline. 

\subsection{Therapeutic-Loaded Sutures for Wound Healing}

Therapeutic sutures are a compelling localized-delivery medium due to their ability to target payloads directly at the wound site while maintaining the conventional surgical procedure. Polymer choice (e.g., PLA/PLGA/PCL families), suture self-assembly structure (mono- vs.\ multi-filament), and method of production (coating, electrospinning, melt processing) appear to be key factors driving efficiency and rate of payload release \citep{deng2021drugelutingsuture}. State-of-the-art engineered systems include PLGA and PLLA electrospun sutures that release a local anesthetic over days while retaining usable tensile strength \citep{weldon2012electrospunsutures}, and the electrospun core, ensuring the sustained delivery of small molecules and proteins \citep{padmakumar2016coresheath}. Moreover, polyester-based coatings can enhance release and mechanical durability. For example, antibiotic-loaded PLA suture coatings with PCL appear to modulate release rate and long-lasting strength via blend composition \citep{liu2020plasutures}.

In protein therapeutics, sustained delivery is particularly important due to the instability of growth factors in the wound microenvironment, ultimately benefiting from a protein-retaining carrier. In the case of EGF-family ligands, controlled release from biodegradable carriers has been shown to improve in vivo wound healing, including rhEGF-loaded PLGA--alginate microspheres in diabetic wound models \citep{gainza2013rhegfplga}. Finally, silk fibroin is widely studied as a mechanically strong and biocompatible substrate for wound application, wherein clinical and preclinical evidence supports silk-based biomaterials as wound dressings \citep{kamalathevan2018silksystematic}, and fabrication protocols for silk fibroin biomaterials are well-established \citep{rockwood2011silkprotocols}.

\end{document}